\include{psfig}

\documentstyle[aps,manuscript,epsf]{revtex}

\def\bm{{\bf m}}

\def\vnabla{{\vec \nabla}}
\def\veta{{\vec n}}

\def\vn{{\vec n}}

\title{Towards a Tetravalent Chemistry of Colloids}
\author{David R. Nelson}
\address{Lyman Laboratory of Physics, Harvard University, 
Cambridge, MA 02138}
\date{\today}
\begin{document}
\draft
\maketitle
\widetext
\begin{abstract}
We propose coating spherical particles or droplets with anisotropic
nano-sized objects to allow micron-scale colloids to link or 
functionalize with a four-fold valence, similar to the sp$^3$ 
hybridized chemical bonds  associated with, e.g.,
carbon, silicon and germanium. Candidates for such coatings include
triblock copolymers, gemini lipids, metallic or semiconducting 
nanorods and conventional liquid crystal compounds. We estimate the 
size of the relevant nematic Frank constants, discuss how to obtain
other valences and analyze the thermal distortions of ground
state configurations of defects on the sphere.
\bigskip
\bigskip

\end{abstract}
\narrowtext

Self-assembly or functionalization of  micron-sized 
particles has many potential uses, including particle-based 
bioassays\cite{svo}, catalysis\cite{bran}
and photonic band gap materials\cite{joan}. However, 
attaching a small, predetermined number of chemical groups,
DNA strands or kinesin molecules to considerably larger colloidal spheres
can be an inefficient process, since the number of ligands per sphere
is highly variable. If the ligand concentration in solution is kept very 
low, many spheres will have no ligands at all. Higher ligand
concentrations require tedious sorting methods such as gel
electrophoresis to select the spheres with the correct number
of attached objects\cite{zan}.

Controlled fabrication of {\it tetravalent} colloids would be of 
particular interest. Dense glassy and crystalline colloidal 
arrays (possibly involving DNA linker elements\cite{mirk}) 
usually display the large number of near neighbors $(Z=12-14)$
characteristic of an isotropic pair potential\cite{gass}. We analyze
here a means by which micron-scale colloids could link or be 
functionalized with, say, a {\it four}-fold valence, similar to the sp$^3$
hybridized chemical bonds on an Angstrom scale associated with,
e.g., carbon, silicon and germanium. We suggest in particular coating
spherical colloid particles (or polymerizable liquid droplets) with 
anisotropic objects such as metallic or semiconducting nanorods,
gemini lipids, triblock copolymers or conventional nematogens. If the 
particles are sufficiently anisotropic, a two-dimensional nematic 
liquid crystal phase will appear on the surface. 
The spherical topology then forces four
strongly repulsive disclination defects into the ground state, a 
circumstance which would allow creation of novel tetravalent colloidal
materials with chemical linkers or DNA strands anchored at the defect
cores. In contrast to fcc and bcc colloid arrays, a tetravalent 
colloidal crystal with a diamond lattice structure and appropriate
dielectric contrast is predicted to have a very large photonic 
band gap\cite{ho}. More generally, 
colloidal particles with a 1-, 2-, 3- or 4-fold 
valence would also allow creation of functionalized
micron-sized objects similar to 
the molecules characteristic of organic chemistry, and could be useful
for bioassays and catalysis.

A two-dimensional nematic at a {\it flat} interface is described
by a Frank elastic free energy, namely\cite{degen}
\begin{equation}
F=\frac{1}{2}\int d^2r
[K_1(\vnabla\cdot\veta)^2+K_3(\vnabla\times\veta)^2]
\end{equation}
where $\vn ({\bf r})$ is a slowly varying director field describing the local 
molecular alignment and $K_1$ and
$K_3$ are, respectively, splay and bend elastic constants.
(The twist elastic constant $K_2$ is absent in two dimensions.) 
The effect of thermal fluctuations (which are very strong in $d=2$)
in a sufficiently large system is to renormalize these elastic 
constants to a single common value $K$\cite{nel}. As a first approximation,
we can take  $K\approx\sqrt{(K_1K_2)}$.
The nature of point-like defects in nematic textures depends 
sensitively on the symmetry of the constituent particles. For tilted
molecules (as might be the case in freestanding smectic liquid 
crystal layers\cite{huan} or Langmuir-Blodgett films\cite{knob}), 
the order parameter is a vector representing the projection of the 
molecules onto the plane. The $2\pi$ rotational symmetry leads to 
defects like the vortex shown in Fig. 1a. An unbinding of vortex-antivortex
pairs causes a transition out of the ordered state whenever $K<2k_BT/
\pi$\cite{nelkos}. In contrast, the order parameter for molecules which 
lie down completely on a liquid or solid interface has the higher
symmetry of a {\it headless} arrow. The symmetry under rotations by 
$\pi$ which results would apply to nematically ordered monolayer
phases of, e.g., 
gemini lipids\cite{men}, ABA triblock copolymers\cite{trib} and nanorods
made of gold\cite{niko}, BaCrO$_4$\cite{kim} or CdSe\cite{manna}. In 
this case, it is well known\cite{degen,mermin} that the vortex in 
Fig. 1a is unstable to the two  disclinations shown in Fig. 1b. The 
180 degree rotational symmetry allows these mutually repelling defects to 
separate---these  defects would be bound with a linear 
potential for an order  parameter with a \emph {vector} symmetry.
Disclination unbinding leads to a transition at a temperature which is
four times smaller, i.e., for 
$K<k_BT/2\pi$.
\begin{figure}
\begin{center}
\centerline{\psfig{figure=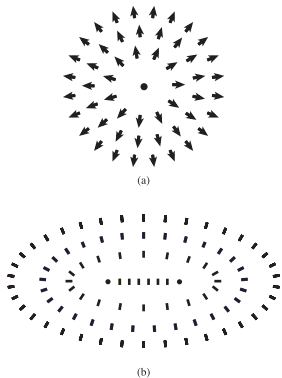}}
\parbox{6.5in}{
\caption{\baselineskip=12pt{\ (a) Vortex configuration of a \emph{vector} order 
parameter in the plane. The order parameter rotates by $2\pi$ on
a circuit around the defect at the center. (b) The texture in 
(a) is unstable to two disclinations 
for a configuration of headless vectors in the plane. The order 
parameter rotates by $\pi$ on a circuit around each of the 
defects marked by small dots.}
}}
\end{center}
\end{figure}

Although pairing of defects and antidefects leads to a virtually
defect-free equilibrium state at high densities or low temperatures 
at flat interfaces, the situation is quite different on surfaces with a 
nonzero Gaussian curvature. For liquid crystal phases embedded in an 
arbitrary curved surface, the free energy in the one Frank constant
approximation reads
\begin{equation}
F=\frac{1}{2} K\int d^2x
\sqrt{g(x)} 
[\partial_i n^j+\Gamma_{ki}^j n^k]^2,
\end{equation}
where $x=(x_1,x_2)$ represents a set of internal coordinates, the 
director is $\vn=\vn(x_1,x_2)$, 
$g$ is the determinant of the metric tensor $g_{ij}(x)$, and the 
$\{\Gamma_{ki}^j\}$ are connection coefficients associated with 
parallel transport of the order parameter. For a rigid 
sphere of radius $R$ with polar coordinates
$(\theta,\phi)$, we have $\sqrt{g}=R^2\sin\theta$, 
$\Gamma_{\phi\phi}^\theta=
-\sin\theta\cos\theta$, $\Gamma_{\theta\phi}^\phi=
\Gamma_{\phi\theta}^\phi=\cot\theta$ and all other
$\Gamma_{ki}^j=0$. The free energy (2) also
describes hexatic liquids (i.e., liquid crystals with a 
{\it six}-fold symmetry) on curved surfaces\cite{nelson,nelpel}. 
In this case, the Gaussian curvature of the sphere produces  an unpairable 
excess of 12 ``disclinations'' (i.e., defects
around which the order parameter rotates by $2\pi/6=60$ degrees) 
at the vertices of an icosahedron in the ground 
state\cite{nelson}. In an elegant paper, Lubensky and Prost have
studied the ground states of more general order parameters with a 
$p$-fold symmetry on the sphere\cite{lub}. In this case, the order
parameter rotates by $2\pi/p$ on a circuit which encloses the minimum 
energy defect, and the $j$-th defect is characterized by an 
integer $n_j=\pm1, \pm 2, \cdots$, which specifies the
winding angle $2\pi n_j/p$.
The Poincar\'e index theorem\cite{spiv} then implies that
\begin{equation}
\sum_j n_j=2p,
\end{equation}
for any ordered texture embedded in a surface with the topology
of sphere. With our conventions, positive winding numbers are
favored by positive Gaussian curvature and negative winding 
numbers are favored by saddle-like regions of negative Gaussian curvature. 

As pointed out by Lubensky and Prost, the 
ground state of a two-dimensional nematic texture 
on a sphere consists of four $n_j=+1$ disclinations at the vertices of a 
tetrahedron\cite{lub}. Aligned states of {\it vector} order 
parameters on the sphere, on the other hand, will be interrupted by
just {\it two} $n_j=+1$ vortices, corresponding to the familiar north 
and south pole singularities of the lines of latitude 
and longitude of spherical cartography. {\it Splay} dominated 
textures for vector and headless vector order 
parameters on the sphere are illustrated in Figs. 2(a)
and 2(b), respectively\cite{bend}. The texture in 
Fig. 2(a) can in fact be transformed into those  of Fig. 2(b) by invoking
the instability shown in Fig. 1. The inevitable defects associated with 
liquid crystalline textures on the sphere could be exploited to
create colloids with a 2- and 4-fold valence in a number of ways.
If, say, gold is deposited on coated solid spheres, the texture will act
as mask such that the two or four ``bald spots'' at the 
defect cores will be coated preferentially. One could then attach, 
e.g., thiol chemical linkers via the extra gold at the bald spots. 
Alternatively, one could introduce DNA linkers attached to 
amphiphillic molecules on polymerizable liquid drops. Linkers 
immiscible with the aligned liquid crystalline molecules
will segregate preferentially at the defect cores, similar to impurity 
segregation at grain boundaries in conventional metallurgy. While we do not wish to minimize the experimental challenges, the 
examples cited above illustrate the possibilities for directional 
bonding associated with ordered states in spheres. \emph{Topology} 
can be used to create  a micron-scale directional chemical bonds 
similar to those produced by quantum mechanics on an Angstrom scale.

\begin{figure}[h]
\begin{center}
\centerline{\psfig{figure=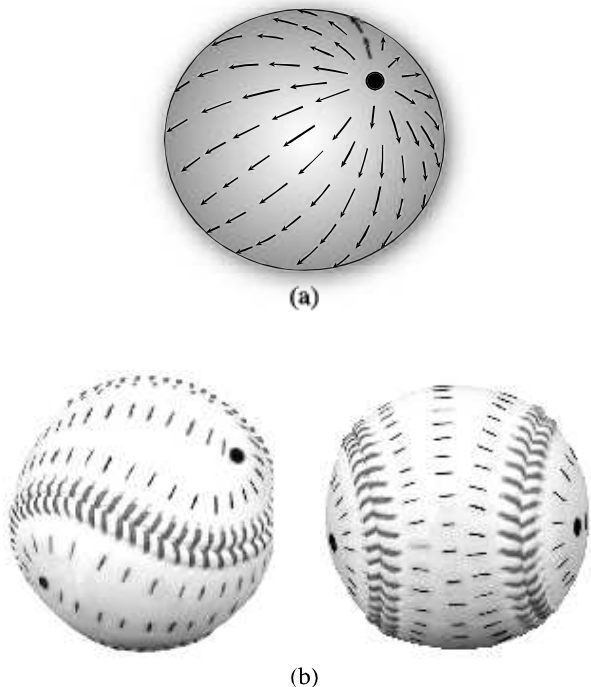}}
\parbox{6.5in}{
\caption{ \ (a) Splay vector order parameter configuration on a sphere.
There are  $+1$ vortices at the north and south pole.
(b) Two views of a splay configuration 
of headless vectors on the sphere. There are four 
disclinations at the vertices of a tetrahedron.
(Note that one row of nematogens tracks  the seam of 
a baseball!)
}}
\end{center}
\end{figure}

To explore the fidelity of directional bonds created by the above
scheme, we need a more detailed analysis of ground states on the sphere 
and their stability to thermal fluctuations. To this end, we exploit
an alternative representation of Eq. (2) in terms of a defect density
\begin{equation}
s(x)=\frac{2\pi}{p}
\sum_j n_j\delta^{(2)}(x-x_j)/\sqrt{g(x_j)}
\end{equation}
and the Gaussian curvature $G(x)$ namely\cite{bow}
\begin{equation}
F=\frac{1}{2} K\int d^2x
\sqrt{g(x)}
\int d^2y\sqrt{g(y)}
[G(x)-s(x)]\frac{1}{\Delta}
\bigg{|}_{_{x,y}}
[G(y)-S(y)].
\end{equation}
Here, $1/\Delta$ denotes the inverse of Laplacian operator,
\begin{equation}
\nabla^2\equiv
\frac{1}{\sqrt{g}}
\sum_j
\partial_i(\sqrt{g} g^{ij}\partial_j), 
\end{equation}
appropriate to a curved surface with metric $g_{ij}(x)$ 
and the defects sit at positions $\{x_j\}$ with a set of integer winding numbers
$\{n_j\}$. The resulting energy for liquid crystal textures with 
a $p$-fold symmetry on the sphere is quite simple,
\begin{equation}
F=-\frac{\pi K}{2p^2}
\sum_{i\not= j} n_in_j
\ln (1-\cos \beta_{ij})+
E(R)\sum_j n_j^2.
\end{equation}
Here the angle $\beta_{ij}=d_{ij}/R$, where $d_{ij}$ is the  geodesic
distance between defects $i$ and $j$, and $E(R)$ is a defect 
self-energy,
\begin{equation}
E(R)=\frac{\pi K}{p^2}
\ln (R/a)+E_c.
\end{equation}
The quantity $E_c$ is a microscopic energy which depends on the details of the particle interactions
near the defect core on length scales of order $a$, where $a$ is the 
particle separation.

The repulsive interaction between defects in (7) is very long-range,
extending out to a significant fraction of the circumference of the sphere. Unlike conventional
Coulomb interactions between charged particles, this interaction
cannot be screened if the order on the surface is strong.
The defect self-energy (8) diverges logarithmically with the sphere radius and 
accounts for the instability indicated
in Fig. 1. To see this, note that the  $2\pi$ vortex of Fig. 1a in a medium
of aligned headless vectors corresponds to a topological charge
$n=2$ with $p=2$. Two such defects at the north and south poles
contribute a logarithmically diverging term in the energy, 
$2[(n/p)^2 \pi K\ln(R/a)]=
2\pi K\ln(R/a)$.  If, however, each defect splits into two disclinations with 
charge $n=1$, the corresponding contribution is $4[(n/p)^2
\pi K\ln(R/a)]=\pi K\ln(R/a)$. We see that the dominant term in the 
energy for large $R/a$ has been reduced by a factor of two. It 
is easy to check that the ground state energy configuration 
with the repulsive interaction in (7) is indeed
two vortices at the north and south poles, and four disclinations at the 
vertices of a tetrahedron for vector and nematic ordering, 
respectively.

The long-range interaction in Eq. (7) will resist thermal disruption
of the ordered defect arrays mentioned above. To investigate this effect for vector order parameters, we 
define a bending angle $\theta$ between the two antipodal defects by
setting  
$\beta_{ij}=\pi+\theta$. Upon expanding (7) in $\theta$, we find that
$F\approx{\rm const.} + (\pi/4) K\theta^2$. The equipartition theorem
now shows that for valence $Z=2$ colloids in the limit
$K>>k_BT$,
\begin{eqnarray}
\langle\cos\beta_i\rangle
&\approx& -1+\frac{1}{2}
\langle\theta^2\rangle\nonumber \\
&\approx& -1+\frac{k_BT}{\pi K}\qquad (Z=2).
\end{eqnarray}
Determining the thermal disruption of the tetrahedral ground state
which results for nematic order on a sphere is more involved.
To get a rough estimate, imagine freezing three of the 
disclinations (with $j=1-3$) at the vertices of a tetrahedron
and then allow the remaining defect (with $j=0$) to fluctuate away from its equilibrium position on the $z$ axis.
Upon expanding Eq. (7) in the polar coordinates $(\theta,\phi)$ of 
this defect about an equilibrium configuration with tetrahedral angles
$\beta_{ij}=\cos^{-1}(-1/3)(i\not= j)$, we obtain $F\approx {\rm const.} 
+(3\pi/16)K\theta^2$, which leads to 
$\langle\theta^2\rangle=8k_BT/3\pi K$. A 
more precise analysis follows from rewriting Eq. (7)
(with $p=2$ and all $n_j=1$) as
\begin{equation}
F=-\frac{\pi K}{8}
\sum_{i\not= j}
\ln(1-\bm_i\cdot\bm_j)
+4E(R),
\end{equation}
where the $\bm_j$ are three-dimensional 
unit vectors specifying the positions of the defects 
on the sphere. Upon writing $\bm_j=\bm_j^0+\delta\bm_j$, where 
$\delta\bm_j\perp\bm_j^0$ represents a small deviation
from a perfect tetrahedral configuration such that 
$\bm_i^0\cdot\bm_j^0=-1/3$ $(i\not= j)$, the energy to quadratic 
order is
\begin{equation}
F= {\rm const.} + \frac{\pi K}{4}
\sum_{i,j} \sum_{\alpha\beta}
M_{ij}^{\alpha\beta}
\delta m_i^\alpha\delta m_j^\beta,
\end{equation}
where $\alpha$ and $\beta$ run over the two independent components of 
the $\{\delta\bm_j\}$ and $M_{ij}^{\alpha\beta}$ is an 
$8\times 8$ matrix describing the deformations of a perfect
tetrahedron on the sphere. The eigenvalues of this matrix 
can be classified according to the irreducible representations 
of the symmetry group of the tetrahedron, similar to the analysis 
of the vibrational modes of the methane molecule\cite{wil}. A
somewhat lengthy calculation shows that the eigenvalue spectrum
of $M_{ij}^{\alpha\beta}$ reads
\begin{equation}
\{ \lambda_j \}=
(3/8)
\{0,0,0,1,1,2,2,2\}.
\end{equation}
Rigid body rotations generate the three normal modes with zero eigenvalues.
The remaining eigenvectors generate two- and three-dimensional
representations of the tetrahedral symmetry group. The doublet 
corresponds to a shear-like twisting deformation of the 
tetrahedron while the non-zero triplet of eigenvalues
represents stretching or compression of vectors
joining neighboring defects. Upon re-expressing the free 
energy (11) in terms of the normal modes we calculate that the 
deviation for valence $Z=4$ colloids from perfect 
tetrahedral angles due to thermal fluctuations is given by
\begin{equation}
\langle\cos\beta_{ij}\rangle
\approx -\frac{1}{3}+
\frac{16}{9\pi}
\frac{k_BT}{K}
\qquad (Z=4).
\end{equation}

Because the distortion from perfect alignment for 
$Z=2$ and $Z=4$ colloids is proportional to $k_BT/K$, 
it is of some interest to determine this quantity for 
a nematic phase formed, say, 
from anisotropic rods [15--17].
Close to the nematic ordering transition, we expect the universal values [9, 12]
$K_1=K_3=K=2k_BT/p^2 \pi$, suggesting fairly large fluctuations 
in the bond angles associated with the topology-induced directional 
bonding for $p=1$ ($Z=2)$ and $p=2$ $(Z=4)$. However, these angles
are more robust to thermal fluctuations in the high density limit where 
$K>>k_BT$. Suppose the rods have length $\ell$ and diameter $d$. 
Estimates are conveniently constructed by adapting results from an 
Onsager theory for the \emph{bulk} Frank constants $K_1^{(3)}$ and $K_3^{(3)}$
in three dimensions, a theory 
valid in the limit $\ell>>d$ \cite{lee}. For a monolayer film of thickness
$d$, we have $K_i\approx dK_i^{(3)}$. For well developed nematic order,
the numerical results of Lee and Meyer \cite{lee}  when applied to thin
films of thickness $d$  lead to the approximate formula
\begin{equation}
K\approx\sqrt{K_1K_3} \approx
0.10 \rho^2\ell^4 k_BT,
\end{equation}
where $\rho$ is the \emph{areal} 
density of nematogens. This density is of order
$\rho\sim 1/\ell^2$ at the ordering transition but rises to 
$\rho\sim 1/\ell d$ deep inside the nematic phase. Thus, in the dense
limit $K=0.10$ $(\ell/d)^2 k_BT$, so that
$k_BT/K\approx 10$ $(d/\ell)^2$. We 
see that the bond angle fluctuations become quite small for large 
$\ell/d$, which is of order 10 in the experiments of Ref. 17.

The Onsager theory calculations of Ref. \cite{lee} show that the bend
elastic constant $K_3$ can be much larger than the splay constant 
$K_1$ for highly anisotropic hard rods. This asymmetry favors the 
``lines of longitude''' texture of Fig. 2a over a texture which 
follows the lines of latitude. These two configurations are 
degenerate in the one Frank constant approximation. A latitudinal
texture would dominate if $K_3<<K_1$. The nematic texture which 
replaces Fig. 2b in this limit is shown schematically in Fig. 3, 
together with possible polymer linking elements emerging from
defects at the vertices of a tetrahedron. Investigations 
of these textures for arbitrary $K_1/K_3$ are currently in progress
using a lattice model of a nematic \cite{vitelli}. Direct simulations
of two-dimensional hard rod fluids \cite{bates} on a sphere would also
be of interest, as would investigations of two-dimensional smectic order 
on a sphere [27].

\begin{figure}[h]
\begin{center}
\centerline{\psfig{figure=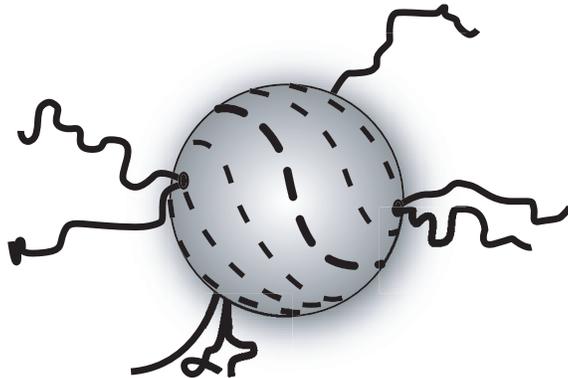}}
\parbox{6.5in}{
\caption{{\baselineskip=12pt \ Bend texture of headless vectors on the sphere appropriate
to $K_1>K_3$. The preferred configuration for $K_1<K_3$ is the splay
texture shown in Fig. (2b). Polymer linkers are shown schematically 
emerging from the tetrahedron of four disclinations.
}}}
\end{center}
\end{figure}

We have argued that coating of colloidal particles with vector or nematic
degrees of freedom could be used to construct micron-sized objects with 
valence $Z=2$ or $Z=4$. Other valences and mechanisms are also possible. For
example, the \emph{bulk} nematic liquid crystal droplets which appear in 
polymer dispersed liquid crystals typically have two surface defects
\cite{mach}, which might also lead to particles with 
functionality $Z=2$. Biaxial nematic droplets should have a single
``boojum'' defect on their surface \cite{mermin}, which upon
polymerization might lead to  a 
$Z=1$ micron-sized analogue of the hydrogen atom. Finally, tetravalent
colloids with, say, DNA linkers could be tethered via one of the four 
defects to a surface (see Fig. 4). This configuration could be useful
for colloid-mediated 
catalysis [2], and the remaining crosslinks in the plane of the 
surface would have a 3-fold valence, i.e. a colloidal analogue of $sp^2$
hybridization.

\begin{figure}[h]
\begin{center}
\centerline{\psfig{figure=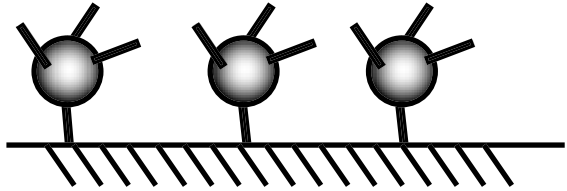}}
\parbox{6.5in}{
\caption{ \ Tetravalent colloids tethered to a planar substrate. The 
remaining linkers can connect with a 3-fold valence $Z=3$.
}}
\end{center}
\end{figure}
\bigskip

\acknowledgements

It is a pleasure to acknowledge helpful 
discussions with P. Alivisatos, S. Block, W. Dressick, F. Dyson,
G. Joyce, M. Kilfoil, R.B. Meyer, C. Mirkin, W. Press, 
V. Vitelli and P. Wiltzius. This work was supported by the NSF
through the Harvard MRSEC via Grant No. DMR98--09363 and through Grant
No. DMR97--14725.

\newpage
\parindent=15pt
\centerline{FIGURES}
FIG. 1. \ (a) Vortex configuration of a \emph{vector} order 
parameter in the plane. The order parameter rotates by $2\pi$ on
a circuit around the defect at the center. (b) The texture in 
(a) is unstable to two disclinations 
for a configuration of headless vectors in the plane. The order 
parameter rotates by $\pi$ on a circuit around each of the 
defects marked by small dots.

FIG. 2. \ (a) Splay vector order parameter configuration on a sphere.
There are  $+1$ vortices at the north and south pole.
(b) Two views of a splay configuration 
of headless vectors on the sphere. There are four 
disclinations at the vertices of a tetrahedron.
(Note that one row of nematogens tracks  the seam of 
a baseball!

FIG. 3. \ Bend texture of headless vectors on the sphere appropriate
to $K_1>K_3$. The preferred configuration for $K_1<K_3$ is the splay
texture shown in Fig. (2b). Polymer linkers are shown schematically 
emerging from the tetrahedron of four disclinations.

FIG. 4. \ Tetravalent colloids tethered to a planar substrate. The 
remaining linkers can connect with a 3-fold valence $Z=3$.
 
\end{document}